\documentstyle[aps,pre,multicol]{revtex}
\begin{document}

\begin{title}
{\bf Soliton motion in a parametrically ac-driven damped Toda lattice}
\end{title}

\author{K.~{\O}. Rasmussen, Boris A. Malomed$^*$, A.~R. Bishop, Niels Gr\o nbech-Jensen}

\address{Theoretical Division and Center for Nonlinear Studies, Los Alamos National Laboratory, Los
Alamos, NM 87545\\
$^*$Department of Interdisciplinary Studies, Faculty of Engineering, Tel
Aviv University, Tel Aviv 69975, Israel}

\maketitle

\begin{abstract}

We demonstrate that a staggered parametric ac
driving term can support stable progressive motion of a soliton in a Toda
lattice with friction, while an unstaggered drivng force cannot. A 
physical context of the model is that of 
a chain of
anharmonically coupled particles adsorbed on a solid surface of a finite
size. The ac driving force models a standing acoustic wave excited on the
surface. Simulations demonstrate that the state left behind the moving
soliton, with the particles shifted from their equilibrium positions,
gradually relaxes back to the equilibrium state that existed before the
passage of the soliton. Perturbation theory
predicts that the ac-driven soliton exists if the amplitude of the drive exceeds
a certain threshold. The analytical prediction for the threshold is in
reasonable agreement with that found numerically. Collisions between two
counter propagating solitons were also simulated, demonstrating
that the collisions are, essentially fully elastic.
\end{abstract}

\begin{multicols}{2}
%\section{INTRODUCTION}

The important role of collective nonlinear excitations in the form of
solitons in several condensed-matter physics contexts 
is widely recognized \cite{general}. Particularly, discrete nonlinear models 
have received an increased amount of attention lately\cite{flach}. Often 
actual observation of nontrivial dynamical effects in realistic systems are
impeded by friction. Compensation of 
friction and support a soliton therefore usually require 
an externally applied driving force \cite{review}. In the simplest case, progressive motion of a soliton
in the presence of dissipation is supported by a dc driving force \cite
{McLSc}, which in fact is the only possibility in a homogeneous continuum
medium. A less obvious but physically important option is
to drive a soliton by an ac force (with no dc component) either in a
continuum system subjected to a periodic spatial modulation \cite{Bob}, or
in a discrete system in the form of a chain of nonlinearly interacting
particles \cite{Finland,angle}.

A simple example of the latter system is the ac-driven damped Toda
lattice (TL) 
\begin{eqnarray}
\ddot{x}_n+\exp \left( x_n-x_{n-1}\right) -\exp \left( x_{n+1}-x_n\right)
=\nonumber \\-\alpha \dot{x}_n+(-1)^n\epsilon \;\sin \left( \omega t\right) \;,
\label{direct}
\end{eqnarray}
where $x_n$ is the displacement of the $n$-th particle, $\epsilon $ and $\omega $ are the amplitude and
frequency of the driving force
and $\alpha $ is the
friction coefficient. This model describes
a chain of nonlinearly coupled particles in a dissipative
environment driven by an externally applied electric  ac field. The alternating sign in
front of the driving term, $\left( -1\right) ^n$ (a discrete pattern of 
this type is frequently 
called {\em staggered }\cite{lanl}), implies alternating electrical charges of the particles in the
chain; without the multiplier $\left( -1\right) ^n$, the driving term can be
trivially eliminated from Eq. (1). In Refs. \cite{Finland}, it was
demonstrated analytically and numerically that the model (\ref{direct})
indeed supports progressive motion of a soliton in the presence 
of finite friction (this was accomplished for this model in the so-called dual
form). Ac-driven motion was predicted and found at the {\em resonant}
soliton velocities 
\begin{equation}
V=\omega /\pi \left( 2\nu +1\right) ,\;\nu =0,\pm 1,\pm 2,...\;.
\label{spectrum_odd}
\end{equation}

The objective of the present work is 
to study, moving solitons in a similar but essentially different model, viz., a {\em
parametrically} ac-driven damped TL, with the driving taken both in the
staggered form, 
\begin{eqnarray}
\ddot{x}_n+\exp \left( x_n-x_{n-1}\right) -\exp \left( x_{n+1}-x_n\right)
=\nonumber \\-\alpha \dot{x}_n+(-1)^n\epsilon \sin \left( \omega t\right) \cdot x_n\;,
\label{parametric}
\end{eqnarray}
and in the unstaggered form, differing from Eq. (\ref{parametric}) by the
absence of the multiplier $(-1)^n$. The model (\ref{parametric}) describes a physical system in
the form, for example, of a chain of particles adsorbed on a solid surface and interacting
through an anharmonic repulsive potential. Additionally, the particles
interact with a uniform external field aligned perpendicular to the
adsorbing surface. (The flield may simply be gravitation, or a dc electric field). In
the latter case, all the particles are assumed to have the same charge (and
the interaction between them within the chain is  generated by the
charges through Coulomb repulsion). The effective drive is induced by a
standing acoustic (elastic) wave excited on the surface, the end particles
in the chain being fixed at the surface edges. If the length of the
surface (i.e. the length of the chain) is $L$, and the number of the
particles is $N+1$, in the equilibrium state the repelling particles are
separated by the distance $L/N$. The finite size of the
surface selects the wavelengths of the standing elastic waves 
$\lambda =2L/M$, where $M$ is an arbitrary integer (the number of the standing half-waves).
Commensurability between the adsorbed chain and the standing wave takes
place provided $M=pN$, where $p=1,2,3,...$ . If the commensurability
index $p$ is odd, starting from $p=1$, the distance between neighboring
particles in the adsorbed chain is equal to an odd number of the standing
half-waves, hence the ac driving forces acting upon the neighboring
particles are $\pi $ out-of-phase, yielding the staggered from of
the driving term in (\ref{parametric}). Oppositely, when $p$ is
even, the phase shift between the neighboring particles is a multiple of 
$2\pi$ corresponding to the unstaggered driving force. Finally, it is necessary
to specify the boundary conditions for the standing elastic wave. If the
edges of the surface are fixed, the particles in the adsorbed chain are
located at the nodes (zero-amplitude points) of the standing wave. It is
easy to see that in this case the external field oriented perpendicular to
the surface gives rise to the direct driving-force term in Eq. (1). On the
other hand, if the edges are free, the particles are located at the
standing-wave maximum-amplitude points, which gives rise to the {\em
parametric} drive in Eq. (\ref{parametric}) (for sufficiently small $x_n$).

In the following we shall first numerically demostrated that the 
model (\ref{parametric}) is able to sustain propagating soliton-like solutions 
which velocity locked to the frequency of the drive. Further, we demonstrate that 
only in the presence of the staggered driving force is the 
soliton like solutions supported.  The locking of the velocity to the driving frequency 
and the 
fact that the soliton like solutions needs a staggered drive to exist 
is then explained analytically and the threshold of the driving amplitude
is found.  Finally the relaxation dynamics of the particles in the wake of the 
soliton is studied analytically.

%\section{NUMERICAL SIMULATIONS}

The system of equations (1) was solved numerically for a chain consisting of 
$500$ particles. The natural initial condition to apply is 
the exact soliton solution of the unperturbed TL
\begin{equation}
x_n\left( t\right) =-\ln \left[ 1+\frac{\left( \xi ^{-2}-1\right)}{ \left( 1+\xi
^{-2\left( n-n_0-Vt\right) }\right)} \right] ,  \label{soliton}
\end{equation}
where the real parameter $\xi$ ($-1< \xi <1$) 
determines the properties of the soliton (width and velocity), and $n_0$ is an arbitrary
phase constant. The velocity $V$ is given as
\begin{equation}
V=\left( \xi -1/\xi \right) /\ln \left( \xi ^{-2}\right).   \label{velocity}
\end{equation}
Launching the soliton (\ref{soliton}) into the system (\ref{parametric})
gives rise to scenarios such as those depicted in Fig. 1(a) and 1(b). In the case 
of Fig. 1(a) the friction is very small ($\alpha=0.01$) and the initial kink 
transforms into a propagating front and in the wake of the front the
particles suffer relatively strong oscillations and only slowly relax to their 
equilibrium position. In the case of stronger friction ($\alpha=0.18$) Fig. 1(b)
shows that the front is still created but the oscillations of the wake particles 
are strongly overdamped and the particles therefore relax rapidly to their initial 
positions. In the latter case the net shift generated by the passing front is zero, which 
in some sense makes the driven front "more solitary" than the soliton of the
unperturbed TL.

The situation is quite different when an unstaggered driving force is applied.
Figure 2 shows a representative example 
of the motion of a driven front in this case. The front is seen to form in much the same fashion as in 
the model with the staggered drive. The front is however decreasing in amplitude 
and eventually ceases to exist. However, it can exist
for rather long times, up to 100 periods ($2\pi/\omega$) of the driving force.

Since the front in the staggered model behaves similarly to a soliton, an interesting experiment is 
to launch counter propagating fronts into the same system and observe their behavior.
An example of such an experiment is shown in Fig. 3 and it is clearly seen that the 
fronts survive the collision. A closer study of trajectories of 
the two fronts shows that the velocity is unchanged by the collision and that 
no phase shift occurs.

%\section{THE\ ANALYTICAL\ APPROACH}
We now investigate the observed phenomenon analytically and 
specify the region of existence in parameter space. We will show that 
the velocity of the propagating front is controlled by the frequency of the driving force.
As in \cite{Finland}, the analytical consideration will be based on
perturbation theory, assuming $\alpha $ and $\epsilon $ to be
sufficiently small. In the zeroth-order approximation, the exact soliton
solution to the unperturbed TL equation is given in Eq. (\ref{soliton}).
Passage of the soliton through a
given particle gives rise to a displacement  
\begin{equation}
\Delta x_n=\int_{-\infty }^{+\infty }\dot{x}_n(t)dt={\mbox{sgn}}\xi \cdot
\ln \left( \xi ^{-2}\right) .  \label{shift}
\end{equation}
If the drive is taken in the staggered form, as in Eq. (\ref{parametric}),
the analysis similar to that of Ref. \cite{Finland} shows that the ac-driven motion of
the soliton is expected at the same resonant velocities (\ref{spectrum_odd}%
). The unstaggered drive, however, gives rise to the resonant velocities 
\begin{equation}
V=\omega /2\pi \nu ,\;\nu =\pm 1,\pm 2,...\;.  \label{spectrum_even}
\end{equation}

The goal of the analytical approach is to predict the minimum ({\em threshold}%
) value $\epsilon _{\mbox{thr}}$ of the ac-drive amplitude $\epsilon $
that is sufficient to compensate the friction and 
allow progressive motion of the soliton. Each value of the integer $\nu $ in (\ref
{spectrum_odd}) and (\ref{spectrum_even}) should thus give rise to a corresponding
function $\epsilon _{\mbox{thr}}\left( \alpha ,\omega \right) $. Steady
motion of the soliton through the damped driven lattice is possible if the
momentum lost by each particle under the action of the friction force, $%
-\alpha \dot{x}_n$, is balanced by the momentum input from the driving
force, $(-1)^n\epsilon \sin \left( \omega t\right) \cdot x_n$. The
momentum loss of the particle can be calculated as 
\begin{equation}
\left( \Delta P\right) _{\mbox{loss}}=-\alpha \int_{-\infty }^{+\infty
}\dot{x}_n(t)dt=-\mbox{sgn}\xi \cdot \alpha \ln \left( \xi ^{-2}\right) ,
\label{loss}
\end{equation}
where Eq. (\ref{shift}) was used. The momentum input is 
\begin{eqnarray}
\left( \Delta P\right) _{\mbox{input}}&=&(-1)^n\epsilon \int_{-\infty
}^{+\infty }{}\sin \left( \omega t\right) \cdot X_n(t)dt\nonumber \\&\equiv&
(-1)^n\epsilon \omega ^{-1}\int_{-\infty }^{+\infty }{}\cos \left( \omega
t\right) \cdot \dot{X}_n(t)dt,  \label{input}
\end{eqnarray}
where integration by parts was performed and the contribution from $t \rightarrow \pm \infty$
was neglected (since it is zero on avarage). The integral in (\ref{input}) can
be calculated explicitly, inserting the expression for $\dot{X}_n(t)$
following from the unperturbed soliton solution (\ref{soliton}).  The result
depends on the arbitrary phase constant $n_0$; the  threshold is determined
by equating the maximum possible absolute value of the momentum input (\ref
{input}) to the absolute value of the momentum loss (\ref{loss}). After a
simple calculation, this leads to the equality 
\begin{equation}
\alpha \ln \left( \xi ^{-2}\right) =\frac{2\pi }\omega \epsilon _{\mbox{thr%
}}\cdot \frac{\left| \sin \left( \frac 12\frac \omega {\xi -1/\xi }\ln
\left( \xi ^{-2}\right) \right) \right| }{\sinh \left( \frac{\pi \omega }{%
\xi -1/\xi }\right) }\;.  \label{balance}
\end{equation}
Replacing the combination $\left( \xi
-1/\xi \right) ^{-1}\ln \left( \xi ^{-2}\right) $ by what follows from Eq. (%
\ref{velocity}), and using the resonant relations
(\ref{spectrum_odd}) or (\ref{spectrum_even}), one 
immediately concludes that in the case (\ref{spectrum_even}), corresponding
to the unstaggered drive, the expression multiplying $\epsilon _{\mbox{thr}%
}$ in Eq. (\ref{balance}) vanishes. This implies  that the threshold ($\epsilon _{\mbox{%
thr}}$) becomes infinite, simply indicating that the progressive motion of the soliton
cannot be supported by the {\em un}staggered ac drive in the presence of
friction, which agrees with our numerical observation. 
However, substitution of Eq. (\ref{spectrum_odd}) for the
staggered driving leads to the meaningful result 
\begin{equation}
\epsilon _{\mbox{thr}}=\frac{\alpha \omega }{2\pi }\ln \left( \xi
_0^{-2}\right) \cdot \sinh \left( \frac{\pi \omega }{\xi _0-1/\xi _0}\right)
\equiv \alpha f\left( \omega \right),  \label{threshold}
\end{equation}
where the parameter $\xi _0$ is determined, as a function of $\nu $ and $%
\omega $, by a transcendental equation following from Eqs. (\ref
{spectrum_odd}) and (\ref{velocity}) 
\begin{equation}
\pi \left( 2\nu +1\right) \left( \xi _0-1/\xi _0\right) =\omega \ln \left(
\xi _0^{-2}\right) .  \label{xi0}
\end{equation}

Note that the proportionality of $\epsilon _{\mbox{thr}}$ to $\alpha $ in
Eq. (\ref{threshold}) is a trivial feature of the first-order approximation
of the perturbation theory \cite{Finland}, while the dependence $\epsilon _{%
{\mbox {thr}}}(\omega )$ is a nontrivial issue. Equations (\ref{threshold}) and (%
\ref{xi0}) give this dependence in an implicit form. To obtain an explicit
dependence, one needs to solve Eq. (\ref{xi0}) for $\xi _0$, which cannot be
done analytically in an exact form, but an approximate solution is
available, 
\begin{equation}
\xi _0^{-1}\approx \frac{2\omega }{\pi \left( 2\nu +1\right) }\ln \left[ 
\frac{2\omega }{\pi \left( 2\nu +1\right) }\right] ,  \label{xi0approx}
\end{equation}
provided that the driving frequency is sufficiently large, so that 
\begin{equation}
\ln \left[ \frac{2\omega }{\pi \left( 2\nu +1\right) }\right] \gg \ln \left[
\ln \left( \frac{2\omega }{\pi \left( 2\nu +1\right) }\right) \right] .
\label{condition}
\end{equation}
In this case, the expression (\ref{threshold}) takes an approximate explicit
form, 
\begin{eqnarray}
\epsilon _{\mbox{thr}}&=&\frac{\alpha \omega }\pi \ln \left( \frac{2\omega }{%
\pi \left( 2\nu +1\right) }\right)\nonumber \\&\times&  \sinh \left[ \frac 12\pi ^2\left(
2\nu +1\right) /\ln \left( \frac{2\omega }{\pi \left( 2\nu +1\right) }%
\right) \right] .  \label{threshold_approx}
\end{eqnarray}

Numerical simulations show that the propagating  soliton can be created and 
sustained at the resonant velocities given by Eq.(\ref{spectrum_odd}) when the
the soliton parameter $\xi$ is chosen in accordance with Eq.(\ref{xi0}).
Further we find that the soliton velocity locks to the driving frequency
even when the initial velocity deviates slightly from the resonate velocity.
That is if the initial velocity 
fulfills $(\omega -\delta \omega)/\pi < V <(\omega +\delta \omega)/\pi$,
where  $\delta \omega \simeq 3$, the soliton will be created and sustained at 
the resonant velocity. 

The
threshold characteristic of the driven motion was measured numerically in
the following way. At a fixed value of the driving amplitude $\epsilon $, the
simulations were run at different values of $\omega $, gradually increasing
the friction coefficient until reaching a maximum value $\alpha _{\max }$,
beyond which the drive could no longer support the stable moving soliton. In
terms of Eq. (\ref{threshold}), 
\begin{equation}
\alpha _{\max }\left( \epsilon ,\omega \right) =\epsilon /f\left( \omega
\right) ,  \label{alpha_max}
\end{equation}
i.e., this procedure implies a way to numerically measure the
function $f(\omega )$.

In Fig.4, the numerically obtained dependence $\alpha _{\max }\left( \omega
\right) $ is plotted, again with the analytically predicted dependence (\ref
{alpha_max}), at the fixed value of the amplitude $\epsilon =0.2$. The
analytical dependence is shown in two forms: the simplified Eq.(\ref
{threshold_approx}) (dashed line), and the more accurate form (dashed-dotted line) 
obtained by numerically
solving the transcendental equation (\ref{xi0}) and substituting the result
into Eq. (\ref{threshold}). The parts of both analytical curves beneath the
turning points are irrelevant: a direct inspection shows that they
correspond either to unphysical solutions with $\xi ^2>1$, or to very broad
solitons, for which, in fact, the model becomes {\em over}damped, and the
perturbation theory does not apply. A conclusion suggested by Fig. 4 is
that, although the accuracy provided by the perturbation theory is not very
high, it captures systematics reasonably. It is also noteworthy that the
simplified approximation (\ref{threshold_approx}) works much worse than the
full perturbative approximation based on Eq. (\ref{xi0}) this is not
surprising: even for the largest frequency, $\omega \equiv 22$, for which
the result of the direct simulations is given in Fig. 1, the condition (\ref
{condition}), necessary for applicability of the simplified approximation,
takes the form $2.63\gg 0.97$.

Finally, the  slow relaxation of the oscillations behind the passing soliton can be
easily analyzed. Indeed, assuming the oscillation amplitude to be small
enough, one can linearize Eq. (\ref{parametric}), yielding
\begin{eqnarray}
\ddot{x}_n -(x_{n-1}&-&2x_n+x_{n+1})=\nonumber \\(&-&1)^n\epsilon \sin \left(
\omega t\right) x_n-\alpha x_n\;.  \label{linear}
\end{eqnarray}
A solution to Eq. (\ref{linear}) can be sought as a sum of a
rapidly oscillating staggered component and a slowly relaxing unstaggered
one 
\begin{eqnarray}
x_n(t)=\left( -1\right) ^nx(t)+X(t).  \label{sum}
\end{eqnarray}
Substituting this into (\ref{linear}) and collecting the staggered and
unstaggered terms, we obtain two simple equations: 
\begin{eqnarray}
\ddot{x}+\alpha \dot{x}+4x&=&\epsilon X\sin \left( \omega t\right),\\
\label{harmonic}
\ddot{X}+\alpha \dot{X}&=&\epsilon \sin \left( \omega t\right) \cdot x.\;
\label{X}
\end{eqnarray}
Using the fact that, in the present approximation, the friction is much
weaker than the inertia, and following the assumption according to which the
unstaggered part of the solution is slowly relaxing (in comparison with the
rapid oscillations of $\sin \left( \omega t\right) $), one can readily
obtain the following approximate solution to Eq. (\ref{harmonic}): 
\begin{equation}
x(t)\approx -\epsilon \left( \omega ^2-4\right) ^{-1}\sin \left( \omega
t\right) \cdot X(t).  \label{x}
\end{equation}
Next, substituting (\ref{x}) into (\ref{X}) and replacing the rapidly 
oscillating term with its avaraged value, we obtain an effective equation governing the slow
relaxation of the state left behind the passing soliton:
\begin{equation}
\ddot{X}+\alpha \dot{X}+\frac 12\epsilon ^2\left( \omega ^2-4\right)
^{-1}X=0.
\label{last}
\end{equation}
Depending on a relation between the
small parameters $\alpha $ and $\epsilon ^2\left( \omega ^2-4\right) ^{-1}$,
this equation may describe both oscillatory and aperiodic relaxation to the final state $X=0$, which is
the equilibrium state that existed before the passage of the soliton. Furthermore,
a numerical study of the relaxations shows that the frequency specified by Eq.(\ref{last}) 
is the relaxation frequency observed in the full simulation.

%\section{CONCLUSION}

In summary we have demonstrated, analytically and numerically, that a staggered
parametric ac driving term can support stable progressive motion of solitons
in a damped Toda lattice, while an unstaggered drive cannot. 
Also, the threshold condition for the
existence of the ac-driven soliton predicted by our perturbation theory is
in reasonable agreement with that found numerically, though there remains a
clear difference. The simulations demonstrate that the state left
behind the moving soliton gradually relaxes back to the equilibrium state
that existed before the passage of the soliton, which we easily
explained analytically. Finally, we demonstrated that collisions between two solitons moving with opposite
velocities are nearly
elastic.

%\section*{ACKNOWLEDGEMENT}

One of the authors (B.A.M.) appreciates visitor support from the Los Alamos
National Laboratory. Work at Los Alamos is performed under the auspices of the US DoE.

\section*{FIGURE CAPTIONS}

Fig. 1:  Typical examples of a moving front with a staggered driving force. (a) is with small
friction ($\alpha =0.01$) and  (b) is 
with relatively strong friction ($\alpha =0.18$). In both cases $\epsilon =0.2$, and $\omega
=12.4$).

Fig. 2: Representative example of a moving front with an unstaggered driving force. The parameters are
$\alpha =0.02$, $\epsilon =0.2$, and $\omega =12.4$).

Fig. 3. Typical example of the quasielastic collision between two
ac-driven solitons. The values of the
parameters are $\epsilon =0.2$, $\alpha =0.1$, and $\omega =12.4$.

Fig. 4. The dependence between the driving frequency $\omega $ and the
maximum friction coefficient $\alpha$ at which a stable moving soliton
can be supported by an ac drive with fixed amplitude $\epsilon =0.2$.
The dots are results of direct numerical simulations 
of Eq. (\ref{parametric}). The dashed-dotted and dashed curves represent, respectively, the
analytical results provided by the full perturbative approximation based on
Eqs. (\ref{threshold}) and (\ref{xi0}), and by the simplified approximation (%
\ref{threshold_approx}). The parts of the curves beneath the turning points
are irrelevant (see text).

\end{multicols}
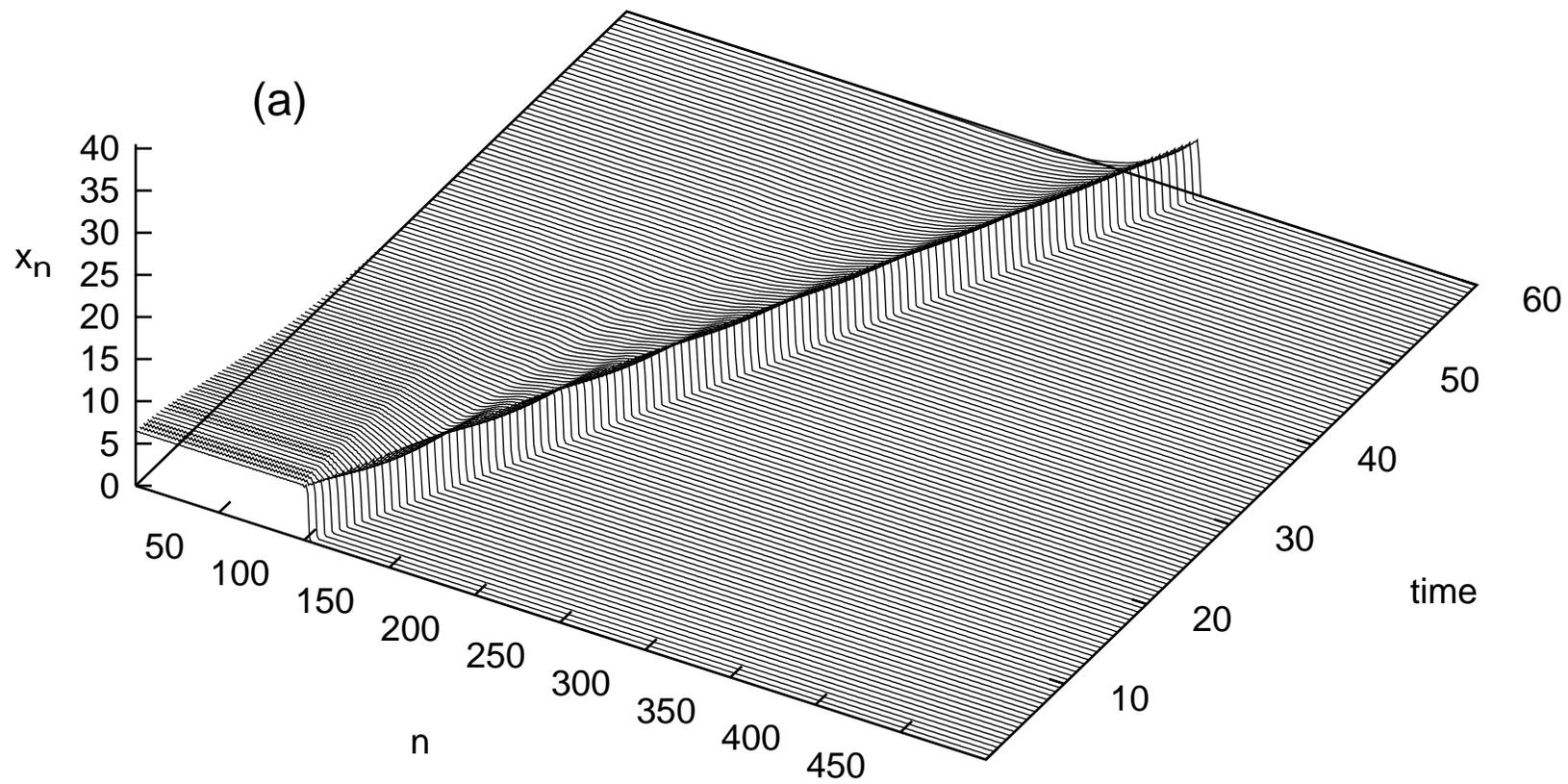
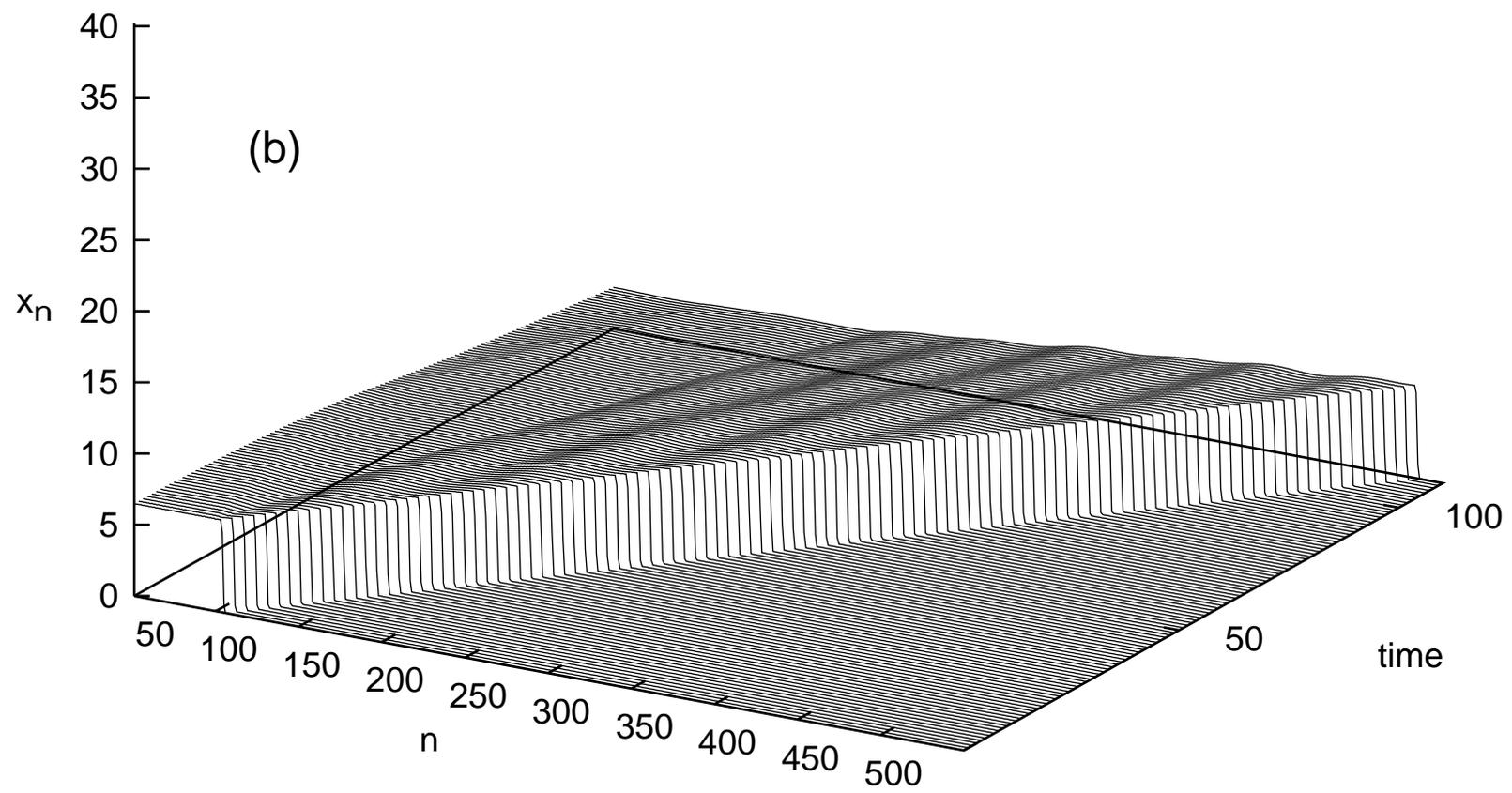
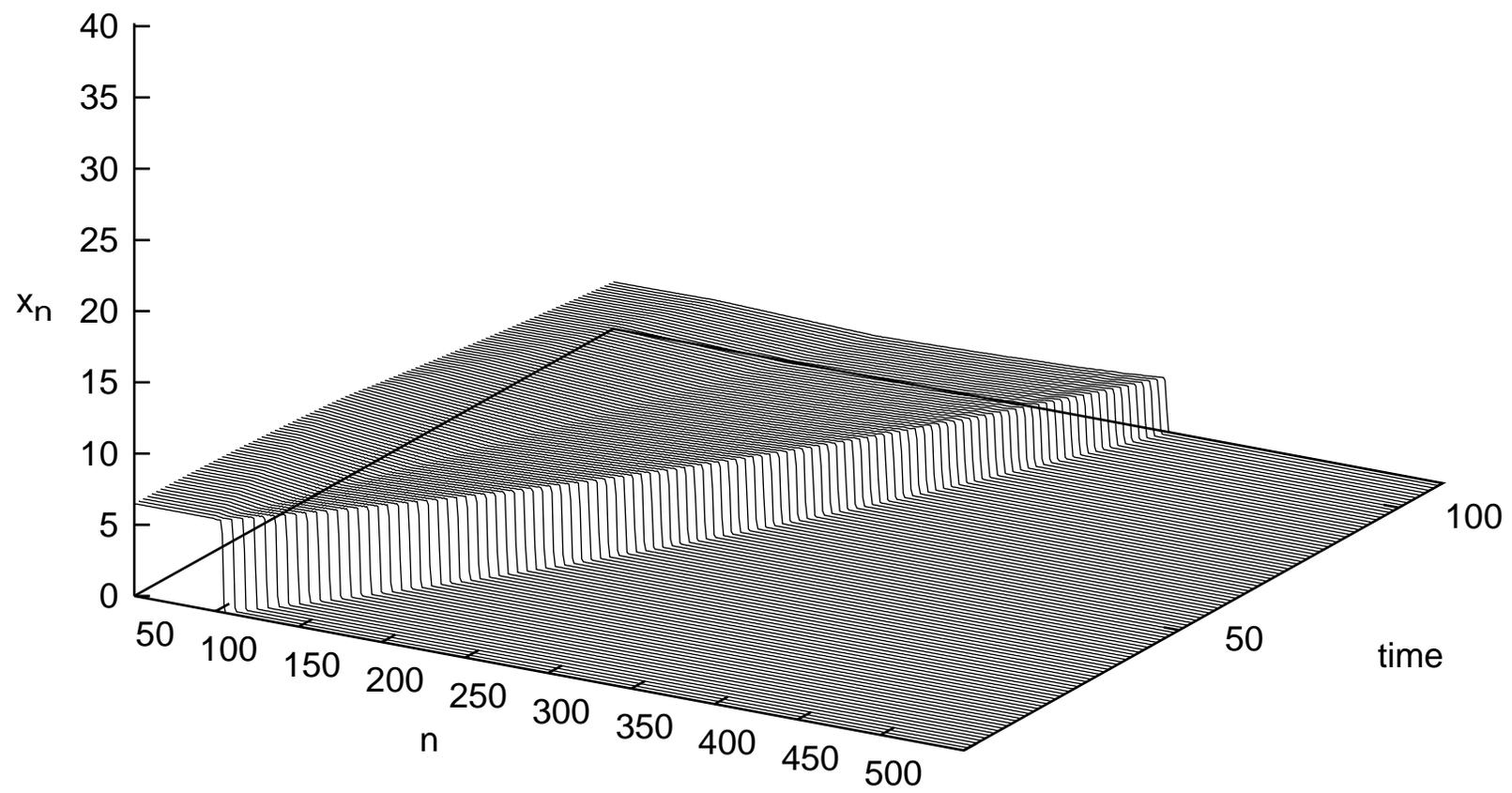
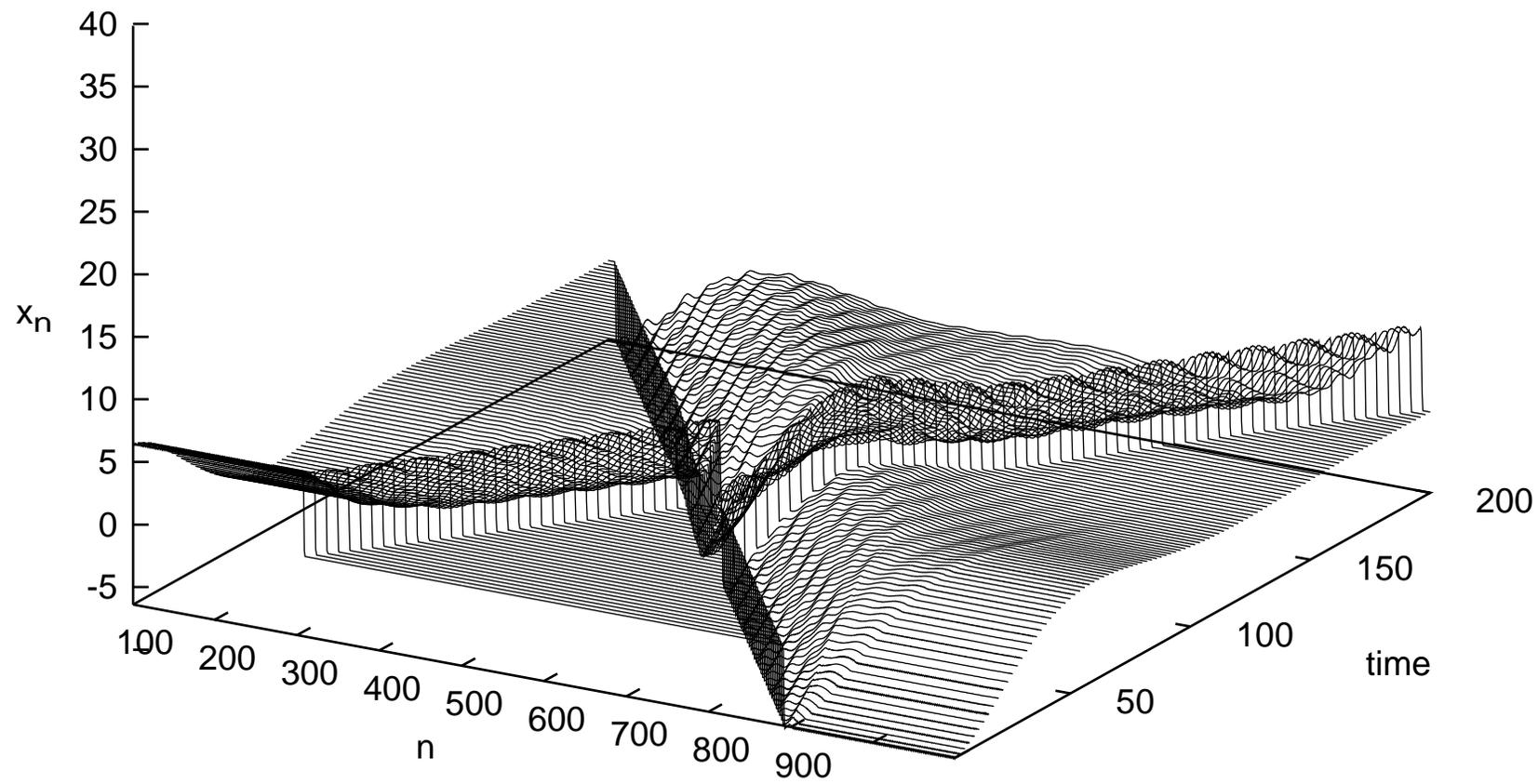
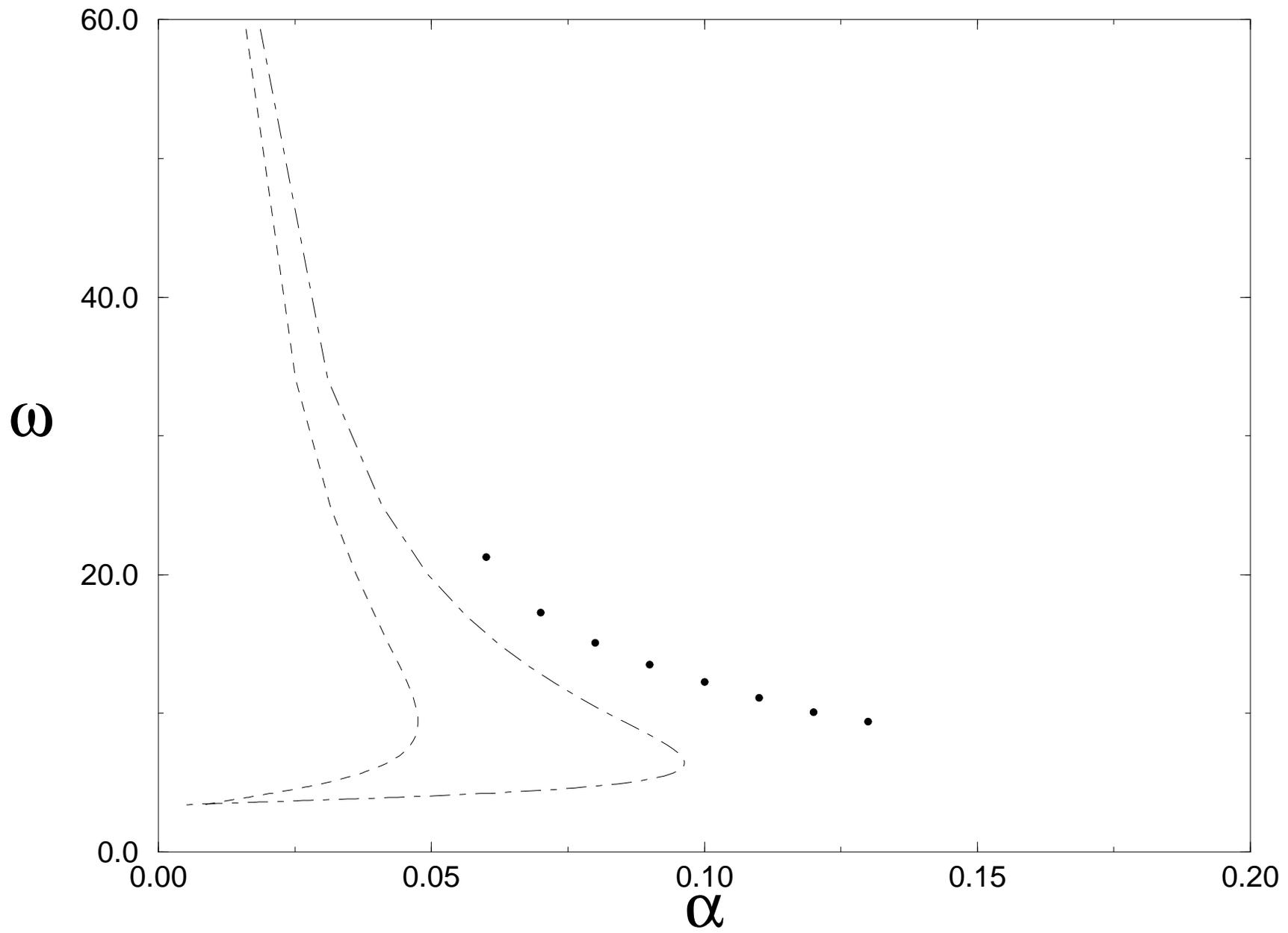
\end{document}